# Fine structure in the energy spectra of ultrasmall Au nanoparticles

Dragomir Davidović* and M. Tinkham

*Department of Physics and Division of Engineering and Applied Sciences, Harvard University, Cambridge, Massachusetts 02138*

We have studied tunneling into individual Au nanoparticles of estimated diameters 2–5 nm, at dilution refrigerator temperatures. The *I-V* curves indicate resonant tunneling via discrete energy levels of the particle. Unlike previously studied normal metal particles of Au and Al, in these samples we find that the lowest energy tunneling resonances are split into clusters of 2–10 subresonances. Such effects appear to be increasingly important in smaller grains, as might be expected from the larger characteristic energies.

In macroscopic metal samples, the typical electron-in-a-box spacing $\delta$ between levels is much less than $k_B T$ at all accessible temperatures. Recently, however, it was demonstrated that, in nm size metallic grains of Al, Au, and Co,[1–3] discrete electronic levels separated by a few meV could be resolved by tunneling spectroscopy at dilution refrigerator temperatures. In this article, we report tunneling data on still smaller grains of gold with diameters 2–5 nm, and containing $\sim 300$–3000 conduction electrons. With these grains, we find that the lowest energy tunneling resonance can be split, and sometimes split into a large number of closely spaced resonances, with a level spacing much smaller than the ''electron-in-a-box'' level spacing $\delta$. The existence of many closely spaced sublevels at the lowest energy level is contrary to the earlier expectations[4] in a spinless model that the spacing between the two lowest energy levels is of order $\delta$. In our further discussion, we refer to this splitting as fine structure.

In this paper we will show the fine structure observed in tunneling spectra of 3 different Au nanoparticles, with estimated diameters $D$ of 2, 3, and 4.7 nm. First we discuss the data.

Figure 1(a) describes our sample geometry. The fabrication of this device is explained in Ref. 2. One nanoparticle is inside the sandwich of $Al_2O_3$, connected by tunneling to two Al leads. In Fig. 1(b) we show a low resolution *I-V* curve of sample 1 at $T=30$ mK. The *I-V* curve is piecewise linear, reflecting the Coulomb staircase. The linear fragments of the curve are much wider in voltage ($\sim 0.2$ V) than are typically found in quantum dots. The large voltage scale arises because the charging energy is large, which indicates that the particle is small in diameter. The junction capacitances and resistances can be obtained by fitting the *I-V* curves using the theory of single electron tunneling:[5] $C_1 + C_2 = 1.4$ aF, $C_1/C_2 = 1.61$, and $R_1 + R_2 = 29$ M$\Omega$. The size of the Coulomb blockade in the *I-V* curve is only $\sim 0.015$ V, because in this sample the background charge $Q_0$ is close to $e/2$, namely $Q_0 = 0.42 e$.[6]

As in our previous work,[2] we will still assume that the particle shape is a hemisphere. Strain and lack of wetting of Au on $Al_2O_3$ makes this assumption unrealistic, but this is not important since we use it only to characterize the particle size from the junction capacitances. The capacitance per unit area in our junctions is $\approx 50$ fF/$\mu$m$^2$.[2] We estimate the particle base diameter $D$ and volume by assuming that the total capacitance of the hemispherical particle is equal to $C_1 + C_2$. The estimated base diameter of sample 1 is 3 nm.

Now we discuss the energy spectrum in this particle. Fig. 1(c) displays the *I-V* curve around the threshold of current conduction, on an expanded scale. The energy level quantization is indicated by the steps in the *I-V* curve. A particle eigenenergy corresponds to a voltage where the current has a step.

In zero magnetic field, every current step in Fig. 1(c) is followed by a region of negative differential conductance. This originates from the BCS density of states in the superconducting aluminum leads.[1] Negative differential resistances can be suppressed by an applied field of 1 T, which destroys the superconductivity in Al. This is shown in Fig. 1(c) in the *I-V* curves taken at 1, 4, and 7 T. We stress that the multiplicity of the observed features is *not* caused by electrostatic jumps or noise, since every step follows a smooth (BCS) line shape vs voltage. Because all features of the *I-V* curve at $H=0$ trace the BCS density of states, and they are equally suppressed by the applied magnetic field, this shows that the resolved steps correspond to eigenstates in which electrons have the same capacitance ratio to the leads, confirming that the eigenstates belong to the same particle.

The estimated electron-in-a-box level spacing, based on particle volume for this particle is few meV. Figure 1(c) shows that there is a much larger number of current steps than expected for this value. The spacing between the resolved low-lying steps is as low as 0.1 meV, after correcting for the capacitive voltage division across the junctions. The current step at the lowest voltage is largest in magnitude. It is followed by a large number of much smaller satellites at higher voltages. At 17.6 mV and 18.5 mV, there are more pronounced increases in current; however, they are not sharp steps. Instead, they are broken into several smaller steps.

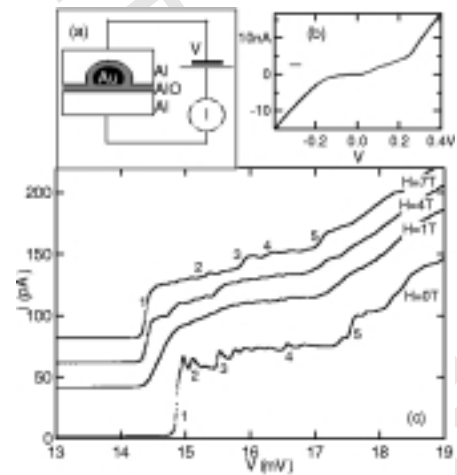

FIG. 1. (a) A device schematic. (b) Large scale *I-V* curve of a 3 nm gold particle (sample 1), at $T=30$ mK. (c) Fine structure *I-V* curve in the 3 nm particle at 30 mK.

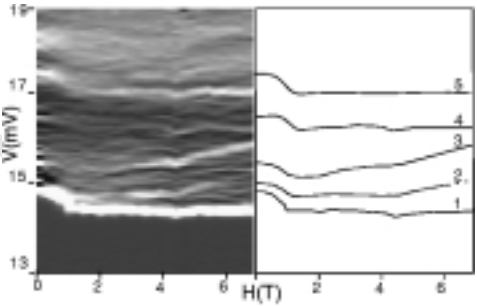

FIG. 2. Magnetic field dependence of the energy spectrum in sample 1. A schematic on the right is a guide to eye, showing magnetic field traces of the most pronounced levels.

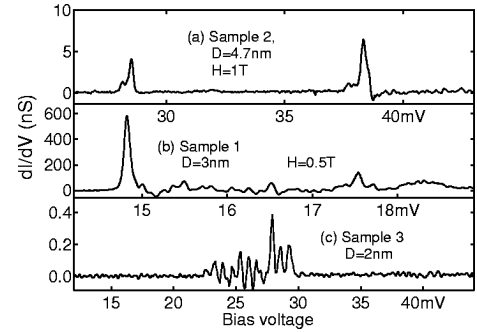

FIG. 3. Fine structure in electronic spectra in three different Au nanoparticles.

In this sample, we can probe the same excitation spectrum at both positive and negative bias. The comparison shows that the fine level structure shown in Fig. 1(c) is also present at the negative bias; however magnitudes of the main and the satellite steps are somewhat different. We also measured this spectrum at $Q_0 = 0.2e$, which showed similar sublevel structure above the larger coulomb blockade voltage.

The magnetic field dependence of the fine structure in sample 1 is shown by the gray scale image of the differential conductance in Fig. 2. The energy levels are much better resolved in zero field, due to the sharpness of the BCS density of states in the leads. The region between 0 and 1 T is dominated by the disappearance of the superconducting gap with field. The most prominent sublevels can easily be traced out in the entire range of magnetic fields. A guide to the eye in Fig. 2(b) depicts the motion of these levels.

In this sample we do not observe simple Zeeman splitting of initially degenerate levels. However, we distinguish two types of level motion with the applied magnetic field. The first type is indicated by traces 1, 2, and 3. The spacing between these levels increases when the magnetic field is increased. The effective g factor corresponding to the relative motion between these levels is $g \sim 0.2 - 0.3$. This value is consistent with the $g$ factors determined by the Zeeman splitting of degenerate levels in larger Au nanoparticles,[2] where $g \sim 0.3$. A second type of level motion is indicated by traces 4 and 5. The spacing between these levels does not change with the applied magnetic field. Small wiggles in the lines at 4.5 T were probably due to a temporary shift of the background charge of the particle.

Before we continue with the interpretation of these results, we show that the fine level structure is also observed in two additional samples. Sample 2 was made with the same fabrication technique as sample 1. Sample 3 was fabricated as described in Ref. 7, and it has chromium leads. Sample 2 has been previously characterized as sample 2 in Ref. 2.

Figure 3 shows the differential conductance vs applied voltage in these 3 different particles above the Coulomb blockade. In samples 1 and 2, a magnetic field is applied only to suppress the superconducting gap in Al. In sample 2 in Fig. 3(a), the 2 lowest energy tunneling resonances are split into 2 resolved subresonances: a main peak, which carries most of the current, preceded by a satellite. The pair of resonances at lower voltage ($\sim 28$ mV) correspond to adding an electron to the particle, and the pair of resonances at higher voltage ($\sim 38$ mV) correspond to removing one. After correcting for the corresponding capacitive division prefactors, the spacing between the two resolved subresonances of each pair is 0.25 meV in energy. In large magnetic field more resolved satellites emerge. At 5 T, careful measurements of the resonance structure show that there are 3 additional weak satellites, following the main peak, with an internal spacing of $\approx 0.2$ meV, in energy. This brings the total number of resolved states contained within a resonance to 5.

In Fig. 3(c), we show the energy spectrum around the ground state of sample 3. The lowest electron-in-a-box level is split into $\approx 10$ maxima, in zero magnetic field. At a voltage of 70 mV (not shown), there is another cluster of resonances. As in sample 2, the resonance clusters are well separated. When the magnetic field is increased to 7 T, the number of sublevels increases by a factor of about 2. However, due to the large number of closely spaced peaks, we were not able to trace out the motion of sublevels with field.

We now return to the interpretation of the data. In the well known constant interaction model, the tunneling resonances occur at energy $E_C + \delta$, where $E_C$ is the charging energy. As the particle diameter $D$ is reduced, both $E_C$ and $\delta$ increase rapidly, as $D^{-2}$ and $D^{-3}$, respectively. The resolution of the tunneling measurements is set by the Fermi distribution in the leads, and it is independent of $D$. Thus, if the splitting of the tunneling resonances of the $I$-$V$ curves or the amplitudes of the satellites are a certain fraction of $E_C$ or $\delta$, then the multiplicity of the levels becomes much more easily observable when the particles are small. We now discuss possible origins of level splitting, and show that they should scale in this way.

First we examine a possibility that the splitting is caused by certain effects outside the grain. When an electron tunnels into the grain, it leads to a stepwise change in the electric field outside the grain. The electric field can induce transitions in the insulator around the particles, because amorphous $Al_2O_3$ might contain charge traps. This process would lead to multiplicity of the tunneling resonances, reflecting the energy spectrum of the charge trap (or several charge traps). It can be shown that the amplitude of the satellites in the tunnel spectrum is proportional to the matrix elements of the electric field between the intitial and the final state of the trap. Thus, as the particle diameter decreases and the electric field increases, the satellite resonances become more pronounced.

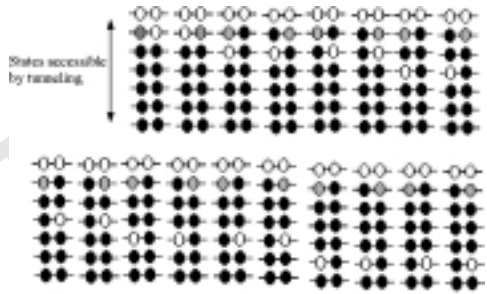

FIG. 4. Electronic configurations of the particle assuming nonequilibrium transport. Number of accessible levels, indicated by arrow, is set by the bias voltage. Black and white circles indicate occupied states and unoccupied states, respectively. Gray circle indicates the state into which an electron just tunneled in.

Splitting of the tunneling resonances can also be caused by electron-electron interactions in the particle. One possibility is that the multiplicity of the observed tunneling resonances is caused by nonequilibrium effects which could occur if the particle has an odd number of electrons (after an electron tunnels in and before an electron tunnels out).[8] This is illustrated in Fig. 4. If the particle relaxation time is much longer than the tunneling time $e/I$, then, after an electron tunnels onto the particle and before an electron exits the particle, the particle can take one of the electronic configurations indicated by Fig. 4. In sample 1, the tunneling rates are much larger than the tunneling rates in previously measured Au particles where nonequilibrium effects have also been observed, suggesting stronger nonequilibrium in this case. If we enumerate all the states that can be created by tunneling in and tunneling out, then the number of states is int($4\,eV_{bias}/\delta$)+2, where $V_{bias}$ is the applied voltage between the two electrodes. Thus, as long as $\delta$ is several times less than $V_{bias}$, we could expect significant clustering near the lowest energy level. Splitting between the satellites is determined by the matrix elements of the electron-electron interaction, and it scales with $\delta$. This could explain why such nonequilibrium states were less pronounced in larger particles, where $\delta$ is smaller. The $g$ factor of $\sim 0.2 - \sim 0.3$ is consistent with $g$ factor of Au particles determined previously.[2]

As an alternative to the nonequilibrium model described above, there is also a possibility that the ground state of the particle has several consecutive Kramers doublets near the Fermi level singly occupied. In that case, the ground state spin of the particle might exceed 1/2. The probabilities of the ground state spin greater than 1/2 for a chaotic, normal metallic particle, have recently been calculated by Brouwer et al.,[9] using a a Hartree-Fock model based on random matrix theory. The theory predicts that in the ground state of a particle, there can be a number of order 1 singly occupied orbitals, and that this number is independent of the particle size, but dependent on the strength of the electron-electron interactions. A nonzero ground state spin has also been predicted by numerical simulations of energy spectra in small chaotic quantum dots,[10] and possible itinerant ferromagnetism[11] in finite size conductors has been proposed. We propose that the fine structure observed in our experiments might reflect the density of many body states near the ground state of such particles. If the SO coupling is strong, then the the crystal field anisotropy lifts the spin degeneracy, except for the overall Kramers degeneracy. In this case, the multiplicity of the tunneling resonances would occur even without nonequilibrium effects. The splitting between the many body states would scale with $\delta$.

To summarize, we have measured that the tunneling resonances in ultrasmall Au nanoparticles are split into several fine sublevels. We divide possible causes of such splitting into the effects outside the particle (charge traps) and electron-electron interactions in the particle (nonequilibrium effects or single occupancy of Kramers doublets). At the present time, it is impossible to narrow down the source of such splitting, but, in any case, the splitting is evidently more pronounced in smaller particles.

We thank Piet Brouwer, Yuval Oreg, Dan Ralph, and Bertrand Halperin for theoretical insights, and S. Shepard for help with fabrication. This work has been supported in part by NSF Grants Nos. DMR-97-01487, DMR-98-09363, PHY-98-71810, and ONR Grant No. N00014-96-0108.

---

*Present address: School of Physics, Georgia Institute of Technology, Atlanta, Georgia 30332.

[1] D. C. Ralph, C. T. Black, and M. Tinkham, Phys. Rev. Lett. **74**, 3241 (1995).

[2] D. Davidović and M. Tinkham, Phys. Rev. Lett. **83**, 1644 (1999).

[3] S. Gueron et al., Phys. Rev. Lett. **83**, 4148 (1999).

[4] O. Agam et al., Phys. Rev. Lett. **78**, 1956 (1997).

[5] D. V. Averin and K. K. Likharev, in *Mesoscopic Phenomena in Solids*, edited by B. L. Altshuler, P. A. Lee, and R. A. Webb (Elsevier, Amsterdam 1991).

[6] In these Au nanoparticles it is extremely difficult to define a lithographic gate, due to the small particle size. The advantage of $Q_0$ being close to $e/2$ is that samples at low bias voltages are found to be much less likely to show current noise due to activated charge traps. To achieve $Q_0 \sim e/2$, we measure several samples, and select a sample with $Q_0 \sim e/2$. At the end of the experiment, we changed $Q_0$ by applying larger voltages, which increased the voltage gap by a factor of $\sim 4$, confirming that the charging energy is indeed large.

[7] D. Davidović and M. Tinkham, Appl. Phys. Lett. **73**, 3959 (1998).

[8] Dan Ralph and Piet Brouwer (private communication).

[9] Piet Brouwer, Yuval Oreg, and B. I. Halperin, Phys. Rev. B **60**, 13 977 (1999).

[10] R. Berkovits, Phys. Rev. Lett. **81**, 2128 (1998).

[11] A. V. Andreev and A. Kamenev, Phys. Rev. Lett. **81**, 3199 (1998).